\documentclass[10pt]{extarticle}
\usepackage{graphicx}
\usepackage{amssymb,amsmath,dsfont}
\def\beq{\begin{eqnarray}}
\def\eeq{\end{eqnarray}}
\usepackage[titletoc,toc,title]{appendix}

\usepackage{amsfonts}

\begin{document}

\title{Spectral sum rules for the Schr\"odinger equation}
\author{Paolo Amore \\
\small Facultad de Ciencias, CUICBAS, Universidad de Colima,\\
\small Bernal D\'{i}az del Castillo 340, Colima, Colima, Mexico \\
\small paolo@ucol.mx}

\maketitle

\begin{abstract}
We study the sum rules of the form $Z(s) = \sum_n E_n^{-s}$, where $E_n$ are the eigenvalues of the 
time--independent Schr\"odinger equation (in one or more dimensions) and $s$ is a rational number 
for which the series converges. We have used perturbation theory to obtain an explicit formula 
for the sum rules up to second order in the perturbation and we have extended it non--perturbatively 
by means of a Pad\'e--approximant. For the special case of a box decorated with one impurity in one 
dimension we have calculated the first few sum rules of integer order exactly; the sum rule of order one
has also been calculated exactly for the problem of a box with two impurities. In two dimensions we have 
considered the case of an impurity distributed on a circle of arbitrary radius and we have calculated
the exact sum rules of order two.
Finally we show that exact sum rules can be obtained, in one dimension, by transforming 
the Schr\"odinger equation into the Helmholtz equation with a suitable density.
\end{abstract}

\section{Introduction}
\label{sec:Intro}

The focus of this paper is on calculating sum rules of the form
\begin{equation}
Z(s) = \sum_{n=1}^\infty \frac{1}{E_n^s} \ ,
\label{eq_sum}
\end{equation}
where $E_n$ are the eigenvalues of the time--independent Schr\"odinger equation for a given Hamiltonian $\hat{H}$ 
and $s$ is a rational number for which the series above converges. 
In particular, Sukumar~\cite{Sukumar90} has discussed sum rules of the kind (\ref{eq_sum}) with integer $s$ for confining 
potentials in one dimension, expressing them directly in terms of integrals of Green's functions.
Crandall~\cite{Crandall96} has studied the case of arbitrary values $s$ by expressing the spectral zeta function 
(\ref{eq_sum}) as a Mellin transform 
\begin{equation}
Z(s) = \frac{i^s}{\Gamma(s)} \int_0^\infty t^{s-1} \int K(x,t|x,0) dx dt
\label{eq_Mellin}
\end{equation}
where $K(x,t|y,0)$ is the spacetime propagator defined as
\begin{equation}
K(x,t|y,0) = \sum_{n=1}^\infty \psi_n(x) \psi_n^\star(y) e^{-i E_n t} \ .
\label{eq_propagator}
\end{equation}

By using the knowledge of the exact propagator (\ref{eq_propagator}) for the case of a perturbed 
oscillator (the isotonic oscillator,  Ref.~\cite{Weissman79}), $V(x)= \frac{\omega^2 x^2}{2} + \frac{g}{x^2}$ (with $g>-1/8$), Crandall was able to 
obtain the expressions  for the corresponding spectral zeta function.
Similarly, he obtained the exact expressions for sum rules of integer order, for problems 
where the Green's function is known explicitly, e.g. for power potentials of the form
$V(x)= |x|^\nu$ (with $\nu>0$).

The main goal of the present paper is to study the case where where neither the propagator nor the Green's 
function are known, but the Hamiltonian can be decomposed in terms of an unperturbed Hamiltonian, for which
both the eigenfunctions and eigenvalues are known, and a perturbation. Extending the approach recently 
introduced in refs.~\cite{Amore19B,Amore19C}, for the case of a heterogeneous Helmholtz equation, we introduce 
Green's functions of rational order and use them to obtain the desired sum rules in terms of suitable traces
involving products of these Green's functions. Unlike in the case discussed in refs.~\cite{Amore19B,Amore19C}, however,
in general the Green's function of order one cannot be calculated explicitly, being the solution of a Schwinger-Dyson
equation, that can be solved perturbatively. Using perturbation theory it is possible to obtain an explicit expression 
for the sum rule of order $s$ (with $s$ rational number for which eq.~(\ref{eq_sum}) converges), even though
the exact eigenvalues and wave functions for the problem are unknown.

For the special cases of impurities in one and two dimensions, for which it is possible to calculate the Green's 
functions exactly, we have derived exact expressions for several sum rules.

The paper is organized as follows: in Section \ref{sec:GF} we discuss the Green's functions of rational order and 
explicitly obtain  their expression up to second order in perturbation theory; in Section \ref{sec:SRrat}
the sum rule of rational order is expressed as a trace of suitable Green's function
up to second order in perturbation theory; in Section \ref{sec:appl} we discuss some applications of the formulas obtained, in one and two dimensions,
comparing the analytical results with purely numerical results. Finally, in Section \ref{sec:concl} we state our conclusions.

\section{Green's functions}
\label{sec:GF}

We consider the Schr\"odinger equation
\begin{equation}
\hat{H} \Psi_n(x) = E_n \Psi_n(x) \ ,
\end{equation}
where $\hat{H} \equiv \hat{H}_0 + \lambda V(x)$ is the total Hamiltonian operator and
$\hat{H}_0$ is the unperturbed Hamiltonian, whose eigenvalues and eigenfunctions are known
\begin{equation}
\hat{H}_0 \psi_n(x) = \epsilon_n \psi_n(x)\ .
\end{equation}
Since our discussion is not limited to one--dimensional problems $n$ it understood to be the set
of quantum numbers that fully identify a quantum state.

The Green's function associated to $\hat{H}_0$ is 
\begin{equation}
G_0(x,y) = \sum_n \frac{\psi_n(x) \psi_n^\star(y)}{\epsilon_n}  \ ,
\end{equation}
since
\begin{equation}
\hat{H}_0 G_0(x,y) = \delta(x-y) \ .
\end{equation}

Unfortunately it is not possible to obtain a closed form for the Green's function associated to 
$\hat{H}$ but one can see that it obeys the Schwinger-Dyson (SD) equation
\begin{equation}
\begin{split}
G(x,y) &= G_0(x,y) - \lambda \int G_0(x,z) V(z) G(z,y) dz \ ,
\label{eq_SchwDys}
\end{split}
\end{equation}
since
\begin{equation}
\begin{split}
\hat{H} G(x,y) &= \hat{H}_0 G_0(x,y)  + \lambda V(x) G_0(x,y) \\
&- \lambda\int \hat{H}_0 G_0(x,z) V(z) G(z,y) dz  + \lambda V(x) \int G_0(x,z) V(z) G(z,y) dz \\
&= \delta(x-y) + \lambda V(x) G_0(x,y) - \lambda V(x) G(x,y) dz  \\
&+ \lambda V(x) \int G_0(x,z) V(z) G(z,y) dz \\
&= \delta(x-y) \ .
\end{split}
\end{equation}

Let us now work out a spectral decomposition of $G$ in the basis of the unperturbed problem
\begin{equation}
\begin{split}
G(x,y) &= \sum_{n,m} Q_{nm} \psi_n(x) \psi_m^\star(y) \ .
\end{split}
\end{equation}

After substituting this equation in the SD equation and projecting over the modes $\psi_a(x)$ and $\psi_b(y)$ we are left with the matrix equation
\begin{equation}
\begin{split}
Q_{ab} &=  \frac{\delta_{ab}}{\epsilon_a} - \lambda \sum_s \frac{\langle a | V | s\rangle}{\epsilon_a} \ Q_{sb} \ .
\end{split}
\end{equation}

Next, by assuming 
\begin{equation}
Q_{ab}= \sum_{j=0}^\infty \lambda^j Q_{ab}^{(j)} \ ,
\label{eq_Q}
\end{equation}
we can write the solution as
\begin{equation}
\begin{split}
Q_{ab}^{(j)} &= \left\{ 
\begin{array}{ccc}
\frac{\delta_{ab}}{\epsilon_a} & , & j=0 \\
- \sum_s  \frac{\langle a | V | s\rangle }{\epsilon_a} \ Q_{sb}^{(j-1)} & , & j>0 \\
 \end{array}
\right.
\end{split}
\end{equation}

The first few corrections take the form
\begin{equation}
\begin{split}
Q_{ab}^{(1)} &= - \frac{\langle a | V | b\rangle}{\epsilon_a \epsilon_b} \\
Q_{ab}^{(2)} &= \sum_{s_1} \frac{\langle a | V | s_1 \rangle \langle s_1 | V | b \rangle}{\epsilon_a \epsilon_{s_1} \epsilon_b } \\
Q_{ab}^{(3)} &= -\sum_{s_1,s_2} \frac{\langle a | V | s_1 \rangle \langle s_1 | V | s_2 \rangle \langle s_2 | V | b \rangle}{\epsilon_a \epsilon_{s_1} \epsilon_{s_2} \epsilon_b } \\
\end{split}
\end{equation}

Formally one can write the solution to all orders as
\begin{equation}
Q_{ab} = \langle a | \left( \mathds{1} - \hat{H}_0^{-1} V + \hat{H}_0^{-1} V \hat{H}_0^{-1} V  -\dots \right) \hat{H}_0^{-1}  | b \rangle
\end{equation}

Following Refs.~\cite{Amore19B,Amore19C} we then define $\tilde{G}_{[1/N]}(x,y)$ satisfying the property
\begin{equation}
\int \tilde{G}_{[1/N]}(x,z_1) \tilde{G}_{[1/N]}(z_1,z_2) \dots \tilde{G}_{[1/N]}(z_N,y)  dz_1 \dots dz_N  = G(x,y)
\label{GFrat_eQ_1}
\end{equation}
with $N \geq 2$.

We can decompose $\tilde{G}_{[1/N]}$ in the basis of the {\sl unperturbed} problem
\begin{equation}
\begin{split}
\tilde{G}_{[1/N]}(x,y) &= \sum_{n,m} q_{nm}^{[1/N]} \psi_n(x)  \psi_m^\star(y) \ ,
\label{GFrat_eQ_2}
\end{split}
\end{equation}
where
\begin{equation}
q_{nm}^{[1/N]} = \int \int \psi_n^\star(x) \tilde{G}_{[1/N]}(x,y) \psi_m(y) dxdy \ .
\end{equation}

We can then  use eq.~(\ref{GFrat_eQ_2}) inside eq.~(\ref{GFrat_eQ_1}) to obtain the  matrix equation
\begin{equation}
\sum_{r_1, \dots, r_N} q_{nr_1}^{[1/N]}  q_{r_1 r_2}^{[1/N]} \dots q_{r_N m}^{[1/N]}  = Q_{nm} \ .
\label{GFrat_eQ_3}
\end{equation}
From this point on we will avoid the superscript $[1/N]$ in the coefficients whenever possible.

Next we express the coefficients as a power series in $\lambda$ as
\begin{equation}
q_{nm} = \sum_{j=0}^\infty q_{nm}^{(j)} \lambda^j \ ,
\label{eq_q}
\end{equation}
and substitute eqs.~(\ref{eq_Q}) and (\ref{eq_q}) inside eq.~(\ref{GFrat_eQ_3}).

To  order $\lambda^0$ we obtain the equation
\begin{equation}
\begin{split}
\sum_{r_1 \dots r_N} q_{nr_1}^{(0)} \dots q_{r_N m}^{(0)} = \frac{\delta_{nm}}{\epsilon_n} \ ,
\end{split}
\end{equation}
whose solution is 
\begin{equation}
q_{nm}^{(0)} = \frac{\delta_{nm}}{\epsilon_n^{1/N}} \ .
\end{equation}

To order $\lambda$, the l.h.s. of eq.~(\ref{GFrat_eQ_3}) takes the form
\begin{equation}
\begin{split}
LHS^{(1)} &= \sum_{r_1 \dots r_N} \left[ q_{nr_1}^{(1)} q_{r_1 r_2}^{(0)} \dots q_{r_N m}^{(0)}  + 
\dots  +  q_{nr_1}^{(0)} q_{r_1 r_2}^{(0)} \dots q_{r_N m}^{(1)}  \right] \\
&= q_{nm}^{(1)}  \sum_{j=0}^{N-1} \frac{1}{\epsilon_n^{j/N} \epsilon_m^{(N-1-j)/N}}  \equiv 
q_{nm}^{(1)}  \eta_{nm}^{[1/N]} \ ,
\end{split}
\end{equation}
and therefore
\begin{equation}
q_{nm}^{(1)} = - \frac{1}{ \eta_{nm}^{[1/N]}} \frac{\langle n | V | m \rangle}{\epsilon_n \epsilon_m} \ .
\end{equation}

To order $\lambda^2$, the l.h.s. of eq.~(\ref{GFrat_eQ_3}) takes the form
\begin{equation}
\begin{split}
LHS^{(2)} &= \sum_{r_1 \dots r_N} \left[ q_{nr_1}^{(2)} q_{r_1 r_2}^{(0)} \dots q_{r_N m}^{(0)}  + 
\dots  +  q_{nr_1}^{(0)} q_{r_1 r_2}^{(0)} \dots q_{r_N m}^{(2)}  \right] \\
&+ \sum_{r_1 \dots r_N} \left[ q_{nr_1}^{(1)} q_{r_1 r_2}^{(1)} q_{r_2 r_3}^{(0)} \dots q_{r_N m}^{(0)} 
+ q_{nr_1}^{(1)} q_{r_1 r_2}^{(0)} q_{r_2 r_3}^{(1)} \dots q_{r_N m}^{(0)} + \dots \right.  \\
& \left.  + 
q_{nr_1}^{(0)} q_{r_1 r_2}^{(0)}  \dots q_{r_{N-1} r_N}^{(1)} q_{r_N m}^{(1)} \right]
\\
&=  q_{nm}^{(2)}  \eta_{nm}^{[1/N]} + \sum_r q_{nr}^{(1)} q_{rm}^{(1)} 
\sum_{j=0}^{N-2} \sum_{l=0}^{N-2-j} \frac{1}{\epsilon_n^{j/N} \epsilon_r^{l/N} \epsilon_m^{(N-2-l-j)/N}} \\
&\equiv  q_{nm}^{(2)}  \eta_{nm}^{[1/N]} + \sum_r q_{nr}^{(1)} q_{rm}^{(1)} \xi_{nrm}^{[1/N]} \ ,
\end{split}
\end{equation}
and therefore
\begin{equation}
q_{nm}^{(2)} =  \frac{1}{ \eta_{nm}^{[1/N]}} \sum_r \frac{\langle n | V | r \rangle \langle r | V | m \rangle}{\epsilon_n \epsilon_r \epsilon_m} \left( 1 - \frac{\xi_{nrm}^{[1/N]}}{\epsilon_r \eta_{nr}^{[1/N]} \eta_{rm}^{[1/N]}}
\right) \ .
\end{equation}

\section{Sum rules of rational order}
\label{sec:SRrat}

The results obtained in the previous section allow us to derive an explicit expression for the sum rules of 
rational order.

In particular, the sum rule of order $1 +1/N$ takes the form
\begin{equation}
\begin{split}
Z\left(1+\frac{1}{N}\right) &= \sum_{n,r} Q_{nr} q_{rn}^{[1/N]}\\
&= \sum_{n,r} Q_{nr}^{(0)} q_{rn}^{[1/N] (0)} \\
&+ \lambda \sum_{n,r} \left[ Q_{nr}^{(0)} q_{rn}^{[1/N] (1)} +  Q_{nr}^{(1)} q_{rn}^{[1/N] (0)}  \right] \\
& + \lambda^2 \sum_{n,r}\left[ Q_{nr}^{(1)} q_{rn}^{[1/N] (1)} + Q_{nr}^{(2)} q_{rn}^{[1/N] (0)} +
Q_{nr}^{(0)} q_{rn}^{[1/N] (2)}  \right] + \dots \ .
\end{split}
\end{equation}

With elementary algebra we obtain
\begin{equation}
\begin{split}
Z^{(0)}(1+1/N) &= \sum_n\frac{1}{\epsilon_n^{1+1/N}} \\
Z^{(1)}(1+1/N) &= - \sum_n \left( 1+ \frac{1}{N} \right) \frac{\langle n | V| n \rangle }{\epsilon_n^{2+1/N}} \\
Z^{(2)}(1+1/N) &= - \frac{s}{2} \sum_{r,n} \frac{\epsilon_n^{-2-1/N}-\epsilon_r^{-2-1/N}}{(\epsilon_n-\epsilon_r)} |\langle n | V| r \rangle|^2  \ .
\end{split}
\end{equation}

It is important to observe that the summand in the expression for $Z^{(2)}$ is finite when
$n=r$ and therefore one can split the double series as 
$\sum_{n,r} = \sum_{n=r}  + \sum_{n\neq r}$; after introducing $s =1 +1/N$ one has
\begin{equation}
\begin{split}
Z^{(2)}(s) &=  \frac{s(s+1)}{2} \sum_{n} \frac{|\langle n | V| n \rangle|^2}{\epsilon_n^{2+s}} \\
&- \frac{s}{2} \sum_{r \neq n} \frac{\epsilon_n^{-1-s}-\epsilon_r^{-1-s}}{(\epsilon_n-\epsilon_r)} |\langle n | V| r \rangle|^2  \ .
\end{split}
\end{equation}

However
\begin{equation}
\frac{\epsilon_n^{-1-s}-\epsilon_r^{-1-s}}{\epsilon_n-\epsilon_r} = 
\left(\epsilon_n^{-2-s} + \epsilon_r^{-2-s} \right) + 
\frac{\epsilon _r \epsilon_n^{-s-2}-\epsilon _n \epsilon _r^{-s-2}}{\epsilon_n-\epsilon _r} \ ,
\end{equation}
and
\begin{equation}
\begin{split}
- \frac{s}{2} &\sum_{r \neq n} \frac{\epsilon_n^{-1-s}-\epsilon_r^{-1-s}}{(\epsilon_n-\epsilon_r)} |\langle n | V| r \rangle|^2 
= - s  \sum_{r\neq n}  \epsilon_n^{-2-s} |\langle n | V| r \rangle|^2 \\
& - \frac{s}{2}  \sum_{r\neq n}  \frac{\epsilon _r \epsilon_n^{-s-2}-\epsilon _n \epsilon _r^{-s-2}}{\epsilon_n-\epsilon _r} |\langle n | V| r \rangle|^2 \ .
\end{split} 
\end{equation}

The first contribution can be simplified using the completeness of the basis
\begin{equation}
\begin{split}
- s \sum_{r\neq n}  \epsilon_n^{-2-s} |\langle n | V| r \rangle|^2 & = 
-s  \sum_{n}  \epsilon_n^{-2-s} \langle n | V \left[ \sum_r | r \rangle \langle r | -  |n \rangle \langle n | \right]  V| n \rangle \\
&= -s \sum_{n}  \epsilon_n^{-2-s} \left( \langle n | V^2| n \rangle 
- \langle n | V| n \rangle^2\right) \ .
\end{split}
\end{equation}

Finally, after these manipulations, the second order correction to the sum rule
takes the form
\begin{equation}
\begin{split}
Z^{(2)}(s) &= \frac{s (s+3)}{2} \sum_{n} \frac{|\langle n | V| n \rangle|^2}{\epsilon_n^{2+s}} - s \sum_{n} \frac{\langle n | V^2| n \rangle}{\epsilon_n^{2+s}} \\
&- \frac{s}{2}  \sum_{r\neq n}  \frac{\epsilon _r \epsilon_n^{-s-2}-\epsilon _n \epsilon _r^{-s-2}}{\epsilon_n-\epsilon _r} |\langle n | V| r \rangle|^2 \ ,
\end{split}
\end{equation}
and the sum rule of order $s$ reads
\begin{equation}
\begin{split}
Z(s) &= \sum_n \frac{1}{\epsilon_n^{s}} - \lambda \sum_n s \frac{\langle n | V| n \rangle }{\epsilon_n^{1+s}} \\
&+ \lambda^2  \left[ \frac{s (s+3)}{2} \sum_{n} \frac{|\langle n | V| n \rangle|^2}{\epsilon_n^{2+s}} - s \sum_{n} \frac{\langle n | V^2| n \rangle}{\epsilon_n^{2+s}} \right. \\
&- \left. \frac{s}{2}  \sum_{r\neq n}  \frac{\epsilon _r \epsilon_n^{-s-2}-\epsilon _n \epsilon _r^{-s-2}}{\epsilon_n-\epsilon _r} |\langle n | V| r \rangle|^2 \right]+ \dots \ .
\label{eq_Zeta}
\end{split}
\end{equation}

We can obtain a non--perturbative extension of the expression above by introducing the simple Pad\'e approximant 
\begin{equation}
\begin{split}
Z^{({\rm Pade})}(s) &= \frac{Z^{(0)}(s) + \lambda \frac{(Z^{(1)}(s))^2- Z^{(0)}(s) Z^{(2)}(s)}{Z^{(1)}(s)}}{1- \lambda \frac{Z^{(2)}(s)}{Z^{(1)}(s)}}
\label{eq_Sum_Pade}
\end{split}
\end{equation}
with a pole at $\lambda Z^{(2)}(s) = Z^{(1)}(s)$. A pole in the sum rules naturally occurs when one eigenvalue gets 
arbitrarily close to zero and the sum rules diverges.

\section{Applications}
\label{sec:appl}

In the following we will discuss the application of eq.~(\ref{eq_Zeta}) to the case of a linear potential in a one--dimensional 
box and the calculation of exact sum rules of integer order for special cases where the Green's functions can be known explicitly.



\subsection{Linear potential in a box}
\label{sub:linear}

In this case the unperturbed Hamiltonian is the Hamiltonian of a particle in a one--dimensional box of size $L$ and the
perturbation is represented by the potential
\begin{equation}
\begin{split}
V(x) = \left\{ \begin{array}{ccc}
\kappa x & , & |x| < L/2 \\
\infty & , & |x| \geq L/2 \\
\end{array}
\right.  \ .
\end{split}
\end{equation}

It is convenient to introduce the dimensionless variable $y = x/L$ ($|y| \leq 1/2$) and cast the Schr\"odinger equation into a dimensionless form
as
\begin{equation}
\begin{split}
\left[ - \frac{1}{2} \frac{d^2}{dy^2} +\rho y  \right] \phi_n(y) = \tilde{E}_n   \phi_n(y) \ ,
\label{eq_adim}
\end{split}
\end{equation}
where
\begin{equation}
\begin{split}
\rho \equiv \kappa \frac{M L^3}{\hbar^2} \hspace{1cm} , \hspace{1cm}
\tilde{E}_n \equiv E_n \frac{M L^2}{\hbar^2}  \ ,
\end{split}
\end{equation}
and
\begin{equation}
\begin{split}
\phi_n(y) &\equiv \psi_n(y L) \ .
\end{split}
\end{equation}

From eq.~(\ref{eq_adim}) we see that we can work with the unperturbed problem corresponding to a box of unit length, $L=1$,
and setting $\hbar = M = 1$.

The matrix elements of the potential in the unperturbed basis are then
\begin{equation}
\begin{split}
\langle m | V | n \rangle &= \left\{ 
\begin{array}{ccc}
0  & , & m=n \\
\frac{4 m n \rho \left((-1)^{m+n}-1\right)}{\pi^2 \left(m^2-n^2\right)^2} & , & m \neq n\end{array}
\right. \\
\langle m | V^2 | n \rangle &=  \left\{ 
\begin{array}{ccc}
\frac{\left(\pi ^2 n^2-6\right) \rho ^2}{12 \pi ^2 n^2}  & , & m=n \\
\frac{4 m n \rho ^2 \left((-1)^{m+n}+1\right)}{\pi ^2 (m-n)^2 (m+n)^2} & , & m \neq n\end{array}
\right. \\
\end{split}
\end{equation}

The spectral sum rule will then read
\begin{equation}
\begin{split}
Z(s) &= \left(\frac{\hbar^2}{M L^2} \right)^{-s} \left[2^s \pi ^{-2 s} \left(\zeta (2 s)-\frac{s \rho ^2
	\zeta (2 (2+s))}{3 \pi ^4}+\frac{2 s \rho ^2
	\zeta (2 (3+s))}{\pi ^6}\right) \right. \\
&+ \left. \rho^2 2^{s+6} \pi^{-2 (s+4)} s \sum_{n\neq r}
\frac{\left((-1)^{n+r+1}+1\right)^2 \left(n^4 r^{-2 s-2}-r^4 n^{-2 s-2}\right)}{\left(n^2-r^2\right)^5}
\right] \\
&+ O\left(\rho^4\right) 
\label{eq_spectral}
\end{split}
\end{equation}

To test this result we have applied the Rayleigh-Ritz method with $2000$ eigenfunctions and we have numerically calculated
the lowest eigenvalues of eq.~(\ref{eq_adim}) for $\rho = k/500$, with $k=0,\dots, 500$.
With these eigenvalues we can approximately calculate the sum rules as
\begin{equation}
Z(s) = \left(\frac{\hbar^2}{M L^2} \right)^{-s} \left[\sum_{n=1}^N  \frac{1}{\tilde{E}_n^s} + 
\sum_{n=N+1}^\infty  \frac{1}{\left(\tilde{E}^{(WKB)}_n\right)^s}  \right] \ ,
\end{equation}
where $\tilde{E}_n$ are the numerical eigenvalues calculated using the Rayleigh-Ritz method and 
$\tilde{E}^{(WKB)}_n$ are the approximations obtained using the WKB method
\begin{equation}
\begin{split}
\tilde{E}^{(WKB)}_n &= \frac{\pi^2 (2n+1)^2}{8} + \frac{\rho ^2}{6 (2 \pi  n+\pi )^2} \\ 
&+ 
\frac{2 \rho^4}{9 (2 \pi  n+\pi )^6} +  \frac{8 \rho ^6}{9 (2 \pi  n+\pi )^{10}} + \dots \ .
\end{split}
\end{equation}

In our calculation we have used $N=500$ because typically the accuracy of the RR eigenvalues decreases as $n$ increases.
For a given value of $s$ we have then calculated the numerical sum rules at the different values of $\rho$ mentioned earlier 
and we have used these results to obtain a fit as a function of $\rho$.

For instance, for $s=1$ we have
\begin{equation}
\begin{split}
Z^{(fit)}(1) &= \left(\frac{\hbar^2}{M L^2} \right)^{-1} \left[
0.333333 + 0.0000881836 \rho^2 + 3.25102 \cdot 10^{-8} \rho^4  \right. \\
&+ \left. 
9.99546 \cdot 10^{-10} \rho^6 - 3.80688 \cdot 10^{-10} \rho^8+ \dots
\right]
\end{split}
\end{equation}
which should be compared with the exact result of eq.~(\ref{eq_spectral}), to order $\rho^2$
\begin{equation}
\begin{split}
Z(1) &= \left(\frac{\hbar^2}{M L^2} \right)^{-1} \left[ \frac{1}{3} -\frac{4 \rho ^2}{14175} + 
0.0003703703703 \rho ^2+ \dots
\right] \\
&\approx  \left(\frac{\hbar^2}{M L^2} \right)^{-1} \left[ 0.333333+ 0.0000881834 \rho ^2+ \dots
\right]  \ .
\end{split}
\end{equation}

Similarly, for $s=3/4$ we have
\begin{equation}
\begin{split}
Z^{(fit)}(3/4) &= \left(\frac{\hbar^2}{M L^2} \right)^{-3/4} \left[
0.789007 + 0.0000975667 \rho^2 + 3.09362 \cdot 10^{-8} \rho^4 \right. \\
&+ \left.  1.15761 \cdot 10^{-9} \rho^6 - 4.46889 \cdot 10^{-10} \rho^8+ \dots
\right]  \ ,
\end{split}
\end{equation}
which should be compared with the exact result of eq.~(\ref{eq_spectral}), to order $\rho^2$
\begin{equation}
\begin{split}
Z(3/4) &= \left(\frac{\hbar^2}{M L^2} \right)^{-1} \left[\frac{2^{3/4} \zeta \left(\frac{3}{2}\right)}{\pi^{3/2}}
-\frac{\rho ^2 \left(\pi ^2 \zeta
	\left(\frac{11}{2}\right)-6 \zeta
	\left(\frac{15}{2}\right)\right)}{2 \sqrt[4]{2}
	\pi ^{15/2}} + \dots
\right] \\
&\approx  \left(\frac{\hbar^2}{M L^2} \right)^{-1} \left[ 0.789011+ 0.0000975665  \rho^2 + \dots
\right]  \ .
\end{split}
\end{equation}

In the present case we cannot apply the diagonal Pad\'e approximant of eq.~(\ref{eq_Sum_Pade}) because the spectral sum rule is an even function
of $\rho$.


\subsection{Infinite box decorated with impurities}
\label{sub:impurity}

The unperturbed Hamiltonian is unchanged with respect to the previous example, while the perturbation is 
represented by the potential
\begin{equation}
V(x) =  \kappa \delta(x-a)  \ ,
\label{eq_impurity}
\end{equation}
with $|a|< L/2$.

In this special case it is possible to solve exactly the Schwinger-Dyson equation (\ref{eq_SchwDys}), as noticed
in Refs.~\cite{Crandall96, Glasser15}, and the Green's function takes the form
\begin{equation}
\begin{split}
G(x,y) &= G_0(x,y) - \frac{\kappa G_0(x,a) G_0(a,y)}{1+\kappa G_0(a,a)} \ .
\label{eq_exact}
\end{split}
\end{equation}

As before it is convenient to cast the time--independent Schr\"odinger equation in a dimensionless form as
\begin{equation}
\begin{split}
\left[- \frac{1}{2} \frac{d^2 }{d \bar{y}^2} + \rho \delta(\bar{y}-\bar{a}) \right] \phi_n(\bar{y})= \tilde{E}_n \phi_n(\bar{y}) \ ,
\end{split}
\end{equation}
where the definitions of $\tilde{E}_n$ and $\phi_n(\bar{y})$ are unchanged while
\begin{equation}
\begin{split}
\rho \equiv \kappa \frac{M L}{\hbar^2}  \hspace{1cm} , \hspace{1cm} \bar{q} \equiv q/L \ .
\end{split}
\end{equation}

The spectral sum rules can then be written as
\begin{equation}
\begin{split}
Z(s) = \sum_{n} \frac{1}{E_n^s} = \Gamma^s  \sum_{n} \frac{1}{\tilde{E}_n^s} \equiv \Gamma^s \bar{Z}(s)
\end{split}
\end{equation}
where $\Gamma \equiv M L^2/\hbar^2$ has dimensions of inverse energy.

In the dimensionless form, the unperturbed Green's function takes the form~\cite{Amore13A}
\begin{equation}
\begin{split}
G_0(\bar{x},\bar{y}) = \frac{1}{2} \left[ -(2 \bar{x}-1) (2 \bar{y}+1) \theta (\bar{x}-\bar{y})-(2
\bar{x}+1) (2 \bar{y}-1) \theta (\bar{y}-\bar{x}) \right] \ ,
\end{split}
\end{equation}
and the full Green's function is obtained from eq.~(\ref{eq_exact}).

In terms of this Green's function we can easily obtain the sum rule of order one, {\sl exact to all orders} as
\begin{equation}
\begin{split}
Z(1) &= \frac{M L^2}{\hbar^2}  \int_{-1/2}^{1/2} G(\bar{x},\bar{x}) d\bar{x} \\
&= \frac{M L^2}{\hbar^2} \frac{4+\rho -16 \rho  \bar{a}^4}{6 \left(2+\rho -4 \rho  \bar{a}^2\right)} \ ,
\end{split}
\label{eq_Sum_Rule}
\end{equation}
with a pole at
\begin{equation}
\rho_c= - \frac{2}{1-4 \bar{a}^2} \ ,
\label{eq_rhoc}
\end{equation}
corresponding to a critical coupling
\begin{equation}
\kappa_c = -\frac{2 \hbar^2}{M L (1-4 a^2/L^2)} \ .
\end{equation}

At $\kappa = \kappa_c$ the sum rule diverges because a bound state  with null energy appears; as soon as $\kappa < \kappa_c$ the energy of the bound state decreases and the sum rule passes from being positive to being negative. The absence of further poles signals that the potential supports only a single bound state.
One should also observe that, if we keep the other parameters fixed, but we let $L \rightarrow \infty$, the critical value $\kappa_c$ approaches $0$ from below, consistent with the well known result that the attractive delta potential always supports a bound state regardless of how small is the coupling.

For a system with  $\rho < \rho_c$, the bound state can disappear as a result of any of the following actions: by making the attractive
potential weaker (i.e. making $\kappa$ smaller), by moving the walls of the box closer (i.e. making $L$ smaller) or by moving the
impurity closer to one of the walls.

Additionally we note that at
\begin{equation}
\rho_0 = -\frac{4}{1- 16 \bar{a}^4}
\end{equation}
the sum rule vanishes. This root is lost when $1- 16 \bar{a}^4= 0$, or $\bar{a}= \pm 1/2$. In this case the Dirac delta falls on the border of the box and therefore its effect disappears.

If we solve the time--independent Schr\"odinger equation for this problem exactly we see that the dimensionless eigenvalues are solutions
to the transcendental equation 
\begin{equation}
\rho \ \frac{-\cos \sqrt{2 \tilde{E}} + \cos \left( \bar{a} \sqrt{8\tilde{E}}\right)}{2\tilde{E}}
+   \frac{\sin \sqrt{2\tilde{E}}}{\sqrt{2\tilde{E}}} = 0 \ .
\label{eq_eigen}
\end{equation}

\begin{figure}
	\centering
	\includegraphics[width=0.7\textwidth]{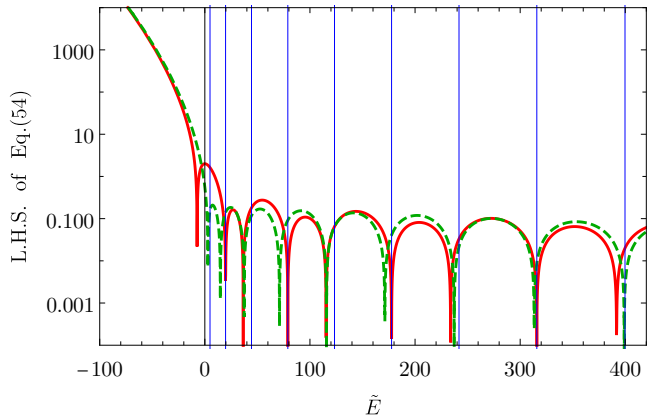}
	\caption{LHS of eq.~(\ref{eq_eigen}) as a function of $\tilde{E}$, 
		for $\rho=-4$; the solid and dashed curves correspond to  $\bar{a} = 0$
		and $\bar{a}=0.4$, respectively. The thin vertical lines correspond
	to the dimensionless eigenvalues of the particle in a box, $\tilde{E} = n^2\pi^2/2$. Observe that for $\bar{a}=0$ the odd states are unaffected by the Dirac delta.}
\label{fig1}
\end{figure}

The critical value of $\rho$ can be obtained from this equation by solving for $\rho$ and letting $\tilde{E} \rightarrow 0$; in this case we obtain eq.~(\ref{eq_rhoc}). The l.h.s. of eq.~(\ref{eq_eigen}) is plotted in Fig.~\ref{fig1} for $\rho=-4$, where  the solid and dashed curves correspond to  $\bar{a} = 0$
and $\bar{a}=0.4$, respectively. 

We have verified eq.~(\ref{eq_Sum_Rule}) by calculating numerically the first $2000$ roots of eq.~(\ref{eq_eigen}) with an accuracy of $100$ digits, for
$\rho=-4$ and $\bar{a}=0$. In this case the sum rule is approximated as
\begin{equation}
\bar{Z}^{({\rm num})}(1) = \sum_{n=1}^{2000} \frac{1}{\tilde{E}^{(num)}_n} \approx  -0.0001013 \ .
\end{equation} 

A much better estimate can be obtained by taking into account the asymptotic behavior of the eigenvalues
\begin{equation}
\begin{split}
\tilde{E}_n &\approx \tilde{E}_n^{\rm (asym)} \equiv \frac{n^2\pi^2}{2} + \sum_{j=1}^\infty c_j(n) \rho^j \ ,
\end{split}
\end{equation}
where
\begin{equation}
\begin{split}
c_1(n) &= 1+(-1)^{1-n} \cos \left(2 n \pi  \bar{a}\right) \\
c_2(n) &= -\frac{3}{4 n^2 \pi^2}+\frac{(-1)^n \cos \left(2 n \pi \bar{a}\right)}{n^2 \pi^2}-\frac{\cos \left(4 n \pi  \bar{a}\right)}{4 n^2\pi ^2} \\
&+\frac{2 (-1)^n \bar{a} \sin \left(2 n \pi  \bar{a}\right)}{n \pi}-\frac{\bar{a} \sin \left(4 n \pi \bar{a}\right)}{n \pi } \\
c_3(n) &=  \frac{5}{2 n^4 \pi^4}-\frac{1}{3 n^2 \pi^2}\\
&+ \frac{(-1)^n \cos \left(2 n \pi  \bar{a}\right) \left(-30+n^2 \pi ^2 \left(3+20 \bar{a}^2\right)\right)}{8 n^4 \pi ^4} \\
&-\frac{\cos \left(4 n \pi  \bar{a}\right) \left(-3+8 n^2 \pi ^2 \bar{a}^2\right)}{2 n^4 \pi^4} \\
&+ \frac{(-1)^n \cos \left(6 n \pi  \bar{a}\right) \left(-6+n^2 \pi ^2 \left(-1+36 \bar{a}^2\right)\right)}{24 n^4 \pi ^4} \\
&-\frac{5 (-1)^n \bar{a} \sin \left(2 n \pi \bar{a}\right)}{n^3 \pi^3}+\frac{4 \bar{a} \sin \left(4 n \pi  \bar{a}\right)}{n^3 \pi^3}+\frac{(-1)^{1+n} \bar{a} \sin \left(6 n \pi  \bar{a}\right)}{n^3 \pi ^3} \\
& \dots \ .
\label{eq_cj}
\end{split}
\end{equation}

Additionally one should observe that for $\bar{a}=0$, the odd eigenfunctions 
are unaffected by the Dirac delta function and therefore 
$\tilde{E}_{2n} = \frac{(2 n)^2 \pi^2}{2}$. In this case we can approximate the sum rule
as
\begin{equation}
\begin{split}
\bar{Z}^{({\rm num})}(1) &= \sum_{n=1}^{1000} \frac{1}{\tilde{E}_{2n-1}} +  \sum_{n=1001}^{\infty} \frac{1}{\tilde{E}^{({\rm asym})}_{2n-1}}  + \sum_{n=1}^{\infty} \frac{1}{\tilde{E}_{2n}} \\
&\approx 1.44 \cdot 10^{-25}  \ ,
\end{split}
\end{equation} 
where we have used the coefficients of eq.~(\ref{eq_cj}).

Let us now modify the unperturbed Hamiltonian of the box with a single impurity by adding a constant term $\gamma$
\begin{equation}
\begin{split}
\hat{H}_0 = - \frac{\hbar^2}{2m} \frac{d^2}{dx^2} + \gamma
\end{split}
\end{equation}

The corresponding Green's function can be easily calculated and it reads
\begin{equation}
\begin{split}
G_0(\bar{x},\bar{y}) &= \frac{{\rm csch}\left(\sqrt{2\gamma}\right)}{\sqrt{2\gamma }}
\left(-\cosh \left(\sqrt{2\gamma} \left(\bar{x}+\bar{y}\right)\right) + \cosh \left(\sqrt{2\gamma } \left(1-|\bar{x}-\bar{y}|\right)\right)\right) \\
\end{split}
\end{equation}

The Green's function for the full problem is obtained as before using eq.~(\ref{eq_exact}). 
By introducing the dimensionless parameter
\begin{equation}
\bar{\gamma} \equiv \frac{\gamma M L^2}{\hbar^2} \nonumber
\end{equation}
we can calculate the sum rule
\begin{equation}
\begin{split}
Z_{\bar{\gamma}}(1) &=  \frac{M L^2}{\hbar^2} \bar{Z}_{\bar{\gamma}}(1) = \frac{M L^2}{\hbar^2} \frac{\mathcal{N}_{\bar{\gamma}}}{\mathcal{D}_{\bar{\gamma}}}
\end{split}
\end{equation}
where
\begin{equation}
\begin{split}
\mathcal{N}_{\bar{\gamma}} &\equiv (-1+\rho ) \sqrt{\bar{\gamma }} 
+\sqrt{2} \coth \left(\sqrt{2\bar{\gamma}}\right) \left(-\rho +\bar{\gamma}\right) \\
&+\rho \ \text{csch}\left(\sqrt{2\bar{\gamma }}\right) 
\left(\sqrt{2} \cosh \left( (2\bar{\gamma})^{3/2}\right)-2 \bar{\gamma }^{3/2} \sinh \left((2\bar{\gamma})^{3/2}\right)\right) \\
\mathcal{D}_{\bar{\gamma}} &\equiv \sqrt{2} \rho  \left(\cosh \left(\sqrt{2\bar{\gamma }}\right)
-\cosh \left((2\bar{\gamma })^{3/2}\right)\right) \text{csch}\left(\sqrt{2\bar{\gamma }}\right) \bar{\gamma}+2 \bar{\gamma }^{3/2}
\end{split}
\end{equation}

\begin{figure}
	\centering
	\includegraphics[width=0.7\textwidth]{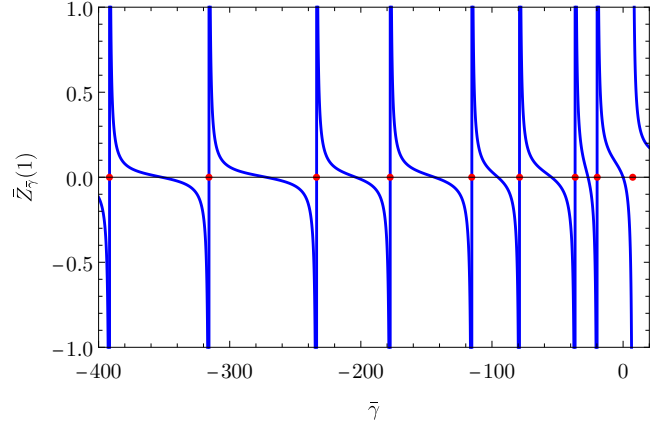}
	\caption{$\bar{Z}_{\bar{\gamma}}(1)$ for $\bar{a}=0$ using  $\rho=-4$ as a function of $\gamma$. The red points are the numerical
		solution of eq.~(\ref{eq_eigen}). }
	\label{fig3}
\end{figure}

As we can see from Fig.~\ref{fig3} the sum rule has poles at $\gamma = - \tilde{\epsilon}_n$, where $\tilde{\epsilon}_n$ is an
exact eigenvalue (in dimensionless form) of the full Hamiltonian. This can be understood since
\begin{equation}
\bar{Z}_{\bar{\gamma}}(1) = \sum_{n=1}^\infty \frac{1}{\tilde{\epsilon}_n + \bar{\gamma}}
\end{equation}

We can use the sum rule above to calculate 
\begin{equation}
\begin{split}
\bar{Z}(j) = \sum_n \frac{1}{\tilde{\epsilon}_n^j}
\end{split}
\end{equation}
as
\begin{equation}
\begin{split}
\bar{Z}(j) = \frac{(-1)^{j-1}}{(j-1)!} \lim_{\bar{\gamma}\rightarrow 0} \frac{d^{j-1}}{d\bar{\gamma}^{j-1}} \bar{Z}_{\bar{\gamma}}(1) 
\end{split}
\end{equation}

In particular
\begin{equation}
\begin{split}
\bar{Z}(2) &= \frac{1}{180 \left(2+\rho -4 \rho  \bar{a}^2\right)^2} 
\left[32 \right. \\
&+ \left. \rho ^2 \left(1+16 \bar{a}^2-160 \bar{a}^4+256 \bar{a}^6+256 \bar{a}^8\right) \right. \\
&+ \left. \rho  \left(8+16 \bar{a}^2 \left(6-40 \bar{a}^2+32 \bar{a}^4\right)\right) \right]  \\
\bar{Z}(3) &= \frac{1}{3780 \left(2+\rho -4 \rho  \bar{a}^2\right)^3} 
\left[256 \right. \\
&+ \left. \rho  \left(78+1440 \bar{a}^2-9408 \bar{a}^4+10752 \bar{a}^6-4608 \bar{a}^8\right) \right. \\
&+ \left. \rho ^2 \left(12+480 \bar{a}^2-3456 \bar{a}^4+27648 \bar{a}^8-24576 \bar{a}^{10}\right) \right. \\
&+ \left. \rho ^3 \left(1+48 \bar{a}^2-432 \bar{a}^4+6912 \bar{a}^8-12288  \bar{a}^{10}-4096\bar{a}^{12}\right)
\right] \\
\bar{Z}(4) &= \frac{1}{226800 \left(2+\rho-4 \rho \bar{a}^2\right)^4} 
\left[ 6144  \right. \\
&+ \left. 64 \rho  \left(37+756 \bar{a}^2-4960 \bar{a}^4+6272 \bar{a}^6-3840 \bar{a}^8+1024 \bar{a}^{10}\right) \right. \\
&+ \left. 16 \rho ^2 \left(-1+4 \bar{a}^2\right)^2 \left(27+1600 \bar{a}^2+4704 \bar{a}^4-10752 \bar{a}^6+5888 \bar{a}^8\right) \right. \\
&+ \left. 48 \rho ^3 \left(-1+4 \bar{a}^2\right)^3 \left(-1-96 \bar{a}^2-672 \bar{a}^4+768 \bar{a}^8\right) \right. \\
&+ 3 \left. \rho ^4 \left(-1+4 \bar{a}^2\right)^4 \left(1+112 \bar{a}^2+1120 \bar{a}^4+1792 \bar{a}^6+256 \bar{a}^8\right)  \right]
\end{split}
\end{equation}

Let us now test the accuracy of eq.~(\ref{eq_Sum_Rule}). As we have seen, when the delta potential becomes sufficiently attractive,
a bound state with arbitrarily small energy appears and the sum rules above diverge. In particular, for $\bar{a}=0$, this corresponds
to $\rho =-2$. 

If we apply the perturbative formulas for $s=1$ and $\bar{a}=0$ the approximate expression for the (dimensionless) sum rule is
\begin{equation}
\bar{Z}(1) =0.333333\, -0.0833333 \lambda  \rho +0.0416456 \lambda ^2 \rho ^2
\end{equation}
and the corresponding Pade approximant of eq.~(\ref{eq_Sum_Pade}) reads
\begin{equation}
\bar{Z}^{\rm Pade}(1) = \frac{0.333333\, +0.0832489 \lambda  \rho }{1+0.499747 \lambda  \rho }
\end{equation}
with a pole at $\rho \approx -2.00101$, remarkably close to the exact value~\footnote{$\lambda$ is just a book--keeping parameter which should be set to $1$ at the end of the calculation.}.
Unfortunately the pole of the Pad\'e approximant moves to $\rho \approx -1.412$ for the sum rule of order $2$, and even worse results are found for $s=3$ and higher. However, the pole of the sum rules, which provides the critical coupling at which a bound state appears, does not 
depend on the order of the sum rule, whereas the pole in the Pad\'e approximant  depends on $s$; it is easy to understand
why the simple Pad\'e of eq.~(\ref{eq_Sum_Pade}) works so well for the sum rule of order one: in that case, the exact sum rule 
takes the form of a diagonal $[1,1]$ Pad\'e, which is precisely the form in  eq.~(\ref{eq_Sum_Pade}). The sum rules of higher orders,
are still diagonal Pad\'e, but of orders $[2,2]$, $[3,3]$, etc.  In those cases, a reliable approximation of the pole would require perturbative calculations of higher order or a different implementation of the Pad\'e approximant.

Let us now consider the case of two Dirac delta located at $x=a$ and $x=b$. 
The time--independent Schr\"odinger equation in dimensionless form will 
read
\begin{equation}
\begin{split}
\left[ - \frac{1}{2} \frac{d^2}{d\bar{y}^2}  + \rho \delta(\bar{y} -\bar{a}) + \mu \delta(\bar{y} -\bar{b}) \right] \phi_n(\bar{y})
= \tilde{E}_n \phi_n(\bar{y})
\end{split}
\end{equation}

The Green's function for the case of two or more impurities can also be constructed explicitly, as recently done in Ref.~\cite{Glasser19}.
The final form is
\begin{equation}
\begin{split}
G(x,y) &=  G_0(x,y) - \frac{1}{D} \left\{
-\rho G_0(x,a) G_0(a,y) - \mu G_0(x,b) G_0(b,y)  \right. \\
&+ \left. \mu \rho \left[ 
 G_0(x,a) \left(G_0(b,y) G_0(a,b) - G_0(b,b) G_0(a,y)\right)  \right. \right.\\
&+ \left. \left. G_0(x,b) \left(G_0(a,y) G_0(b,a) - G_0(a,a) G_0(b,y)\right)
\right]
\right\}
\end{split}
\end{equation}
where
\begin{equation}
\begin{split}
D &\equiv \left( 1 + \rho  G_0(a,a) \right) 
\left( 1 + \mu  G_0(a,a) \right) - \mu \rho  G_0(a,b) G_0(b,a) 
\end{split}
\end{equation}

Our formula does not reproduce completely eq.~(9) of Ref.~\cite{Glasser19}, because of the a sign difference in our definitions of Green's function (which amounts to change the couplings $(\rho,\mu) \rightarrow (-\rho,-\mu)$) and in two typos in the formula in Ref.~\cite{Glasser19}. 

Also in this case we can easily calculate the sum rule of order one to all orders
\begin{equation}
\begin{split}
Z(1) &= \frac{M L^2}{\hbar^2} \left[
\mathcal{F}(\bar{a},\bar{b},\rho,\mu) \theta(a-b) + \mathcal{F}(\bar{b},\bar{a},\mu,\rho) \theta(-a+b) \right]
\label{eq_sumruledelta2}
\end{split}
\end{equation}
where
\begin{equation}
\begin{split}
\mathcal{F}(a,b,\rho,\mu) &= \frac{\mathcal{N}(\bar{a},\bar{b},\rho,\mu) }{\mathcal{D}(\bar{a},\bar{b},\rho,\mu) }
\end{split}
\end{equation}
and
\begin{equation}
\begin{split}
\mathcal{N}(\bar{a},\bar{b},\rho,\mu)  &\equiv 
4+\rho  \left(1-16 \bar{a}^4\right)+\mu \left(1-16 \bar{b}^4\right) \\
&-2 \mu  \rho  \left(-1+2 \bar{a}\right) \left(\bar{a}-\bar{b}\right) \left(1+2 \bar{b}\right)  \\
&\cdot \left(1+4 (\bar{a}^2+ \bar{b}^2- \bar{a} \bar{b} )+2 (\bar{b}-\bar{a}) \right) \\
\mathcal{D}(a,b,\rho,\mu)  &\equiv 12+6 \rho  \left(1-4 \bar{a}^2\right)
+6 \mu  \left(1-4
\bar{b}^2\right) \\
&-12 \mu 
\rho  \left(-1+2 \bar{a}\right)
\left(\bar{a}-\bar{b}\right) \left(1+2
\bar{b}\right) \\
\end{split}
\end{equation}

\begin{figure}
	\centering
	\includegraphics[width=0.47\textwidth]{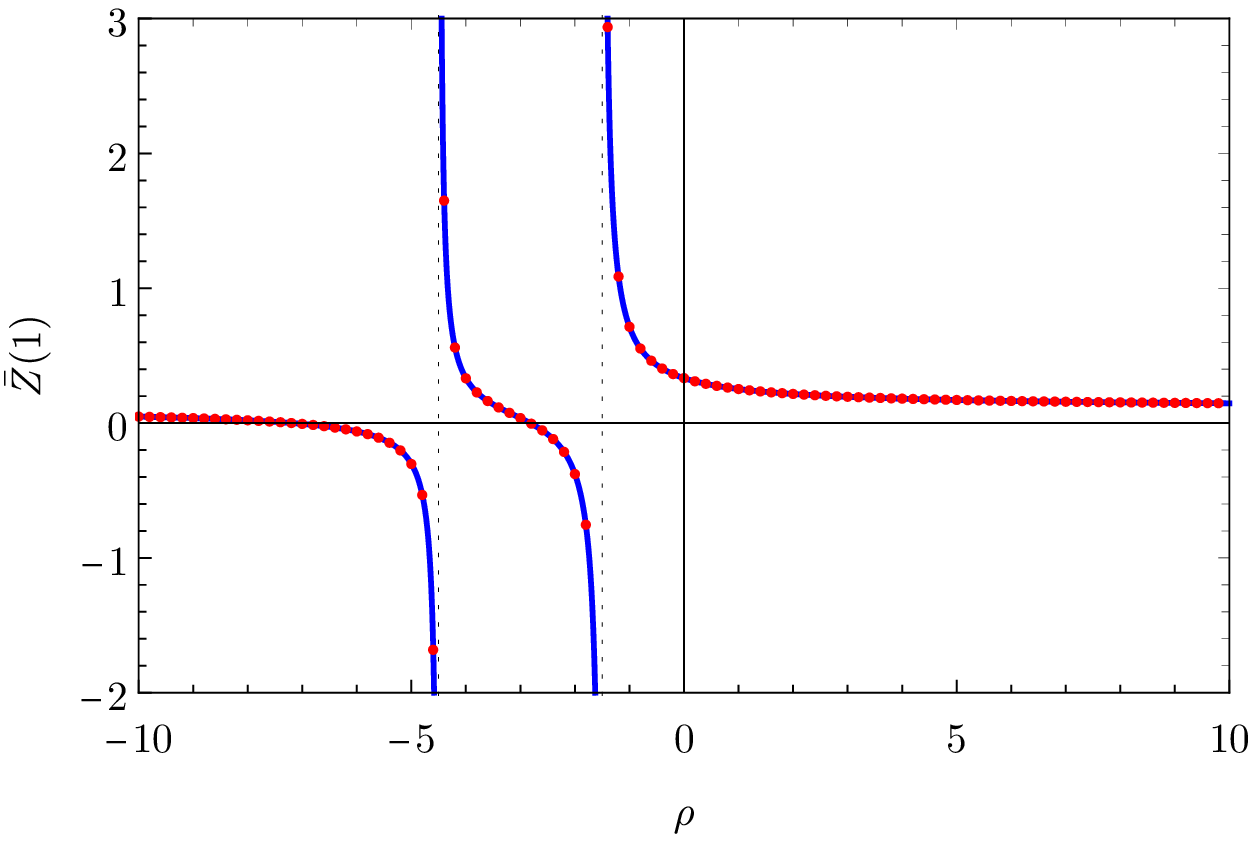} \hspace{.3cm} 	\includegraphics[width=0.47\textwidth]{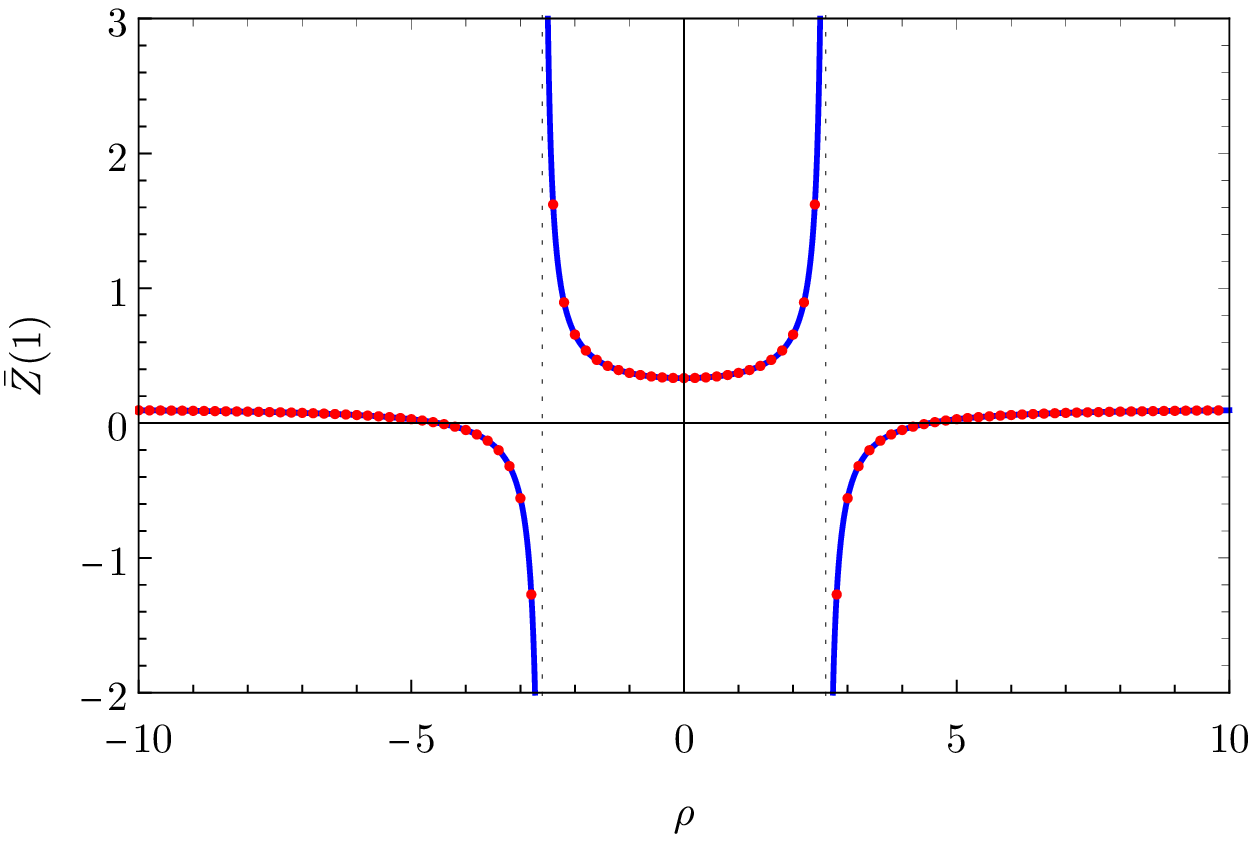}
	\caption{$\bar{Z}(1)$ for $\bar{a}=-\bar{b} = 1/6$ using  $\mu=\rho$ (left plot) and
		$\mu=-\rho$ (right plot). The solid lines are the exact results of eq.~(\ref{eq_sumruledelta2}), whereas the dots
	are the numerical results obtained with the Rayleigh-Ritz method with $2000$ basis elements. }
	\label{fig2}
\end{figure}

Notice that for $\bar{b} = \bar{a}$, eq.~(\ref{eq_sumruledelta2}) correctly reduces to eq.~(\ref{eq_Sum_Rule}) for a potential $(\rho + \mu) \delta(x-\bar{a})$.

In Figure \ref{fig2} we plot $\bar{Z}(1)$ for $\bar{a}=-\bar{b} = 1/6$ using  $\mu=\rho$ (left plot) and
$\mu=-\rho$ (right plot).  The solid lines are the exact results of eq.~(\ref{eq_sumruledelta2}), whereas the dots
are the numerical results obtained with the Rayleigh-Ritz method with $2000$ basis elements. For the case of the right plot,
which corresponds to $\mu = -\rho$ and $\bar{b}=-\bar{a}$, the sum rule is an even function of $\rho$ and there is only one bound
state (the two singularities merely correspond to configurations that are one the reflection of the other).
For $\mu=\rho$, one the other hand, the sum rule has two singularities, located at $\rho_1 = -(1-2\bar{a})^{-1}$
and $\rho_2 = - (2 \bar{a} (1-2 \bar{a}))^{-1}$: the second singularity corresponds to the critical value of $\rho$ at which
a second bound state appears.

\subsection{Two dimensional regions decorated with impurities}

Consider a circle of radius $R$ and Dirichlet boundary conditions at the border; the Green's function obeys
the equation
\begin{equation}
\begin{split}
- \frac{\hbar^2}{2 M} \Delta G_0 - E_\gamma G_0 = \delta^{(2)}({\bf r} - {\bf r}')
\end{split}
\end{equation}

As usual it is convenient to write this equation in a dimensionless form by introducing the definitions:
\begin{equation}
\begin{split}
\bar{r} &\equiv \frac{r}{R} \\
\bar{\gamma} &\equiv \frac{2 M R^2 E_\gamma}{\hbar^2}\\
\bar{G}_0 &\equiv \frac{\hbar^2}{2 M R^2} G_0
\end{split}
\end{equation}
and the corresponding equation becomes
\begin{equation}
\begin{split}
-  \Delta \bar{G}_0 - \bar{\gamma} \bar{G}_0 = \delta^{(2)}(\bar{\bf r} - \bar{\bf r}')
\end{split}
\end{equation}

The Green's function satisfying the equation above is~\cite{Duffy}
\begin{equation}
\begin{split}
\bar{G}_0(\bar{r},\theta,\bar{r}',\theta') &=  -\frac{1}{4} \sum_{n=-\infty}^\infty \cos \left[n (\theta - \theta')\right] \\
&\times J_n(\bar{k}_0 \bar{r}_<) \left[ Y_n(\bar{k}_0 \bar{r}_>)- \frac{Y_n(\bar{k}_0)}{J_n(\bar{k}_0 )} J_n(\bar{k}_0 \bar{r}_>)
\right]
\end{split}
\end{equation}
where
\begin{equation}
\begin{split}
\bar{k}_0 \equiv \sqrt{\bar{\gamma}} \hspace{0.5cm} , \hspace{0.5cm}
\bar{r}_> \equiv \max (\bar{r},\bar{r}') \hspace{0.5cm} , \hspace{0.5cm}
\bar{r}_< \equiv \min (\bar{r},\bar{r}') \nonumber \ .
\end{split}
\end{equation}

We can cast the Green's function in a more compact form as
\begin{equation}
\begin{split}
\bar{G}_0(\bar{r},\theta,\bar{r}',\theta') &=  \sum_{n=0}^\infty g_n(\bar{r},\bar{r}') \cos \left[n (\theta - \theta')\right]
\end{split}
\end{equation}
where
\begin{equation}
\begin{split}
\bar{g}_0(\bar{r},\bar{r}') &=  -\frac{1}{4}  J_0(\bar{k}_0 \bar{r}_<) \left[ Y_0(\bar{k}_0 \bar{r}_>)
- \frac{Y_0(\bar{k}_0)}{J_0(\bar{k}_0 )} J_0(\bar{k}_0 \bar{r}_>)\right] \\
\bar{g}_n(\bar{r},\bar{r}') &= -\frac{1}{2}  J_n(\bar{k}_0 \bar{r}_<) \left[ Y_n(\bar{k}_0 \bar{r}_>)- \frac{Y_n(\bar{k}_0)}{J_n(\bar{k}_0 )} J_n(\bar{k}_0 \bar{r}_>) \right] \hspace{0.5cm} , \hspace{0.5cm} n \geq 1 \\
\end{split}
\end{equation}

We may be tempted to generalize our previous discussion for the one--dimensional delta function to the two--dimensional case, 
but in doing so we would immediately stumble into a problem: the formal solution for the Green's function in eq.~(\ref{eq_exact}) 
is spoiled by the short distance behavior of the two--dimensional unperturbed Green's function
\begin{equation}
\lim_{\vec{\eta} \rightarrow 0} G_0(\vec{r},\vec{r}+\vec{\eta}) = \infty
\end{equation}

The two-dimensional Dirac delta potential is an example of quantum mechanical problem where renormalization is needed~\cite{Mead91,Tarrach91,Jackiw91,Bender99,Holstein14}.

Following Ref.~\cite{Mead91} we consider the potential
\begin{equation}
\begin{split}
V(r) = \kappa \frac{\delta(r-r_0)}{2\pi r}
\end{split}
\end{equation}
and the corresponding Hamiltonian
\begin{equation}
\begin{split}
\hat{H} = - \frac{\hbar^2}{2 M} \Delta + \kappa \frac{\delta(r-r_0)}{2\pi r}
\end{split}
\end{equation}

The Green's function associated to this Hamiltonian obeys the equation
\begin{equation}
\begin{split}
-  \Delta \bar{G} - \bar{\gamma} \bar{G} + \frac{\rho}{2 \pi \bar{r}} \delta(\bar{r} - \bar{r}_0) \bar{G} = \delta^{(2)}({\bf r} - {\bf r}')
\end{split}
\end{equation}
where $\rho \equiv 2 M \kappa/\hbar$.

The Dyson-Schwinger equation for the Green's function can be expressed in terms of the integrals
\begin{equation}
\begin{split}
\mathcal{I}_{k+1}(\bar{r},\theta,\bar{r}',\theta') &\equiv - \int_0^1 \int_0^{2\pi} \mathcal{I}_{k}(\bar{r},\theta,\bar{r}_1,\theta_1) \bar{V}(\bar{r}_1) \bar{G}_0(\bar{r}_1,\theta_1,\bar{r}',\theta') r_1 dr_1 d\theta_1 \\
\end{split}
\end{equation}
with
\begin{equation}
\begin{split}
\mathcal{I}_1(\bar{r},\theta,\bar{r}',\theta') &\equiv \int_0^1 \int_0^{2\pi} \bar{G}_0(\bar{r},\theta,\bar{r}_1,\theta_1) \bar{V}(\bar{r}_1) \bar{G}_0(\bar{r}_1,\theta_1,\bar{r}',\theta') r_1 dr_1 d\theta_1 \\
&= \frac{\rho}{2} \sum_{n=0}^\infty (1+\delta_{n0}) \cos \left[n (\theta - \theta')\right] g_n(\bar{r},\bar{r}_0) g_n(\bar{r}_0,\bar{r}') \\
\end{split}
\end{equation}

It is easy to see that 
\begin{equation}
\begin{split}
\mathcal{I}_k(\bar{r},\theta,\bar{r}',\theta') &= (-1)^{k+1} \left(\frac{\rho}{2}\right)^k \\
&\times \sum_{n=0}^\infty (1+\delta_{n0})^k
 \cos \left[n (\theta - \theta')\right] g_n(\bar{r},\bar{r}_0) \left[ g_n(\bar{r}_0,\bar{r}_0) \right]^{k-1} g_n(\bar{r}_0,\bar{r}') \nonumber
\end{split}
\end{equation}

Finally, the full Green's function reads
\begin{equation}
\begin{split}
\bar{G}(\bar{r},\theta,\bar{r}',\theta') &= \bar{G}_0(\bar{r},\theta,\bar{r}',\theta') - \sum_{k=1}^\infty \mathcal{I}_k(\bar{r},\theta,\bar{r}',\theta')  \\
&= \sum_{n=0}^\infty \tilde{g}_n(\bar{r},\bar{r}') \cos \left[n (\theta - \theta')\right] 
\end{split}
\end{equation}
where
\begin{equation}
\begin{split}
\tilde{g}_n(\bar{r},\bar{r}')  &=  \left\{ g_n(\bar{r},\bar{r}')   - \left(\frac{\rho}{2}\right)  
\frac{ (1+\delta_{n0}) g_n(\bar{r},\bar{r}_0)  g_n(\bar{r}_0,\bar{r}')}{1 +\frac{\rho}{2} (1+\delta_{n0}) g_n(\bar{r}_0,\bar{r}_0)} \right\} \\
\end{split}
\end{equation}

The sum rule of order $2$ can be calculated by extracting the zero--energy Green's function of order two,  
expanding the energy dependent Green's function in $\bar{\gamma}$ and selecting the linear contribution, 
which is then used to  calculate the corresponding trace:
\begin{equation}
\begin{split}
Z(2) &= \sum_{n=0}^\infty z_n(2) 
\end{split}
\end{equation}
where
\begin{equation}
\begin{split}
z_0(2) &= \frac{1}{64
	\left(\rho  \log \left(r_0\right)-2 \pi \right)^2} 
\left[ 4 \rho ^2 \left(r_0^2-1\right){}^2-2 \pi \rho  \left(11 r_0^4-16 r_0^2+5\right) \right. \\
&+ \left. \rho \log \left(r_0\right) \left(5 \rho +r_0^4
	\left(-5 \rho +4 \rho  \log
	\left(r_0\right)+24 \pi \right)+2 \rho  \log
	\left(r_0\right)-8 \pi \right) \right. \\
&+ \left. 8 \pi ^2 \right] \\
z_1(2) &= \frac{1}{96 \left(\rho -\rho  r_0^2+4 \pi \right)^2} \left[ (\rho +4 \pi )^2
   -16 \rho  (\rho +4 \pi ) r_0^2 \right. \\
&+ \left. \rho  r_0^4 \left(31 \rho +2 r_0^2
	\left(-9 \rho +\rho  r_0^2+20 \pi \right) \right. \right. \\
&+ \left. \left. 48
	\log \left(r_0\right) \left(\rho  \log
	\left(r_0\right)-4 \pi \right)+16 \pi
	\right) \right] \\
z_2(2) &= \frac{1}{288
	\left(\rho -\rho  r_0^4+8 \pi \right)^2} \left[ (\rho +8 \pi )^2 \right. \\
&+ \left. \rho  r_0^4 \left(64
	(\rho +6 \pi )+2 \rho  r_0^8-128 (\rho +4 \pi
	) r_0^2 \right. \right. \\
	&+ \left. \left. r_0^4 \left(61 \rho -72 \rho  \log
	\left(r_0\right)+112 \pi \right)+72 (\rho +8
	\pi ) \log \left(r_0\right)\right) \right] \\
	&\dots 
\end{split}
\end{equation}

A direct inspection of the above expressions shows the presence of poles at
\begin{equation}
\begin{split}
\rho_c^{(0)} &= \frac{2\pi}{\log r_0} \\
\rho_c^{(j)} &= \frac{4\pi j}{1- r_0^{2j}} \hspace{0.5cm} , \hspace{0.5cm} j=1,2,\dots \\
\end{split}
\end{equation}
corresponding to the critical couplings at which one of the eigenvalues vanishes. 

We can easily verify this result for the ground state by writing explicitly the zero energy wave function
\begin{equation}
\begin{split}
\Psi_0(r) &= N \left[ - \theta(r-r_0) \log r  - \theta(r_0-r)  \log r_0 \right]
\end{split}
\end{equation}
where $N$ is the normalization constant (irrelevant for our calculation).

By substituting this expression inside the Schrodinger equation we obtain
\begin{equation}
\begin{split}
-\Delta \Psi_0(r) + \frac{\rho}{2 \pi r} \delta (r-r_0) \Psi_0(r) = \frac{\delta(r-r_0)}{r_0} \left( 1 - \frac{\log r_0  \ \rho}{2\pi}\right) = 0
\end{split}
\end{equation}
whose solution provides the critical coupling $\rho_c^{(0)}$.

Similarly, for the states with non--vanishing angular momentum ($m>0$), the zero--energy solutions are
\begin{equation}
\begin{split}
\Psi_m(r) &= N_m \left[  \theta(r-r_0) \sinh (m \log r) + \theta(r_0-r)  \frac{1}{2} r^m \left(1-r_0^{-2
	m}\right) \right]
\end{split}
\end{equation}
and the corresponding Schrodinger equation is
\begin{equation}
\begin{split}
-\Delta \Psi_m(r) &+ \frac{m^2}{r^2} \Psi_m(r) + \frac{\rho}{2 \pi r} \delta (r-r_0) \Psi_m(r) \\
&= \frac{{r_0}^{-m-1} \delta (r-{r_0}) \left(\rho  \left({r_0}^{2m}-1\right)-4 \pi  m\right)}{4 \pi } = 0
\end{split}
\end{equation}
whose solution provides the critical coupling $\rho_c^{(m)}$.

 For $0 < r_0 \ll 1$ and keeping $\rho$ fixed, the sum rule behaves as
\begin{equation}
\begin{split}
Z(2)  &= \left( \frac{1}{32} + \frac{5 \rho -8 \pi }{64 \rho  \log ({r_0})} + \dots \right) + 
\left( \frac{1}{96}-\frac{7 \rho {r_0}^2}{48 (\rho+4 \pi )} + \dots \right) \\
&+ \left( \frac{1}{288} + \dots \right) + \left( \frac{1}{640} + \dots \right) + \left( \frac{1}{1200} + \dots \right) + \dots
\end{split}
\end{equation}
where the leading contributions have the general form $\frac{(1-\delta_{n0}/2)}{8 (n+1)^2 (n+2)}$
(see also eq. (5.18) of Ref.~~\cite{Steiner87}). 

Notice that 
\begin{equation}
\sum_{n=0}^\infty \frac{(1-\delta_{n0}/2)}{8 (n+1)^2 (n+2)} =  \frac{\pi^2}{48} - \frac{5}{32}
\end{equation}
is the sum rule of order two for the unit circle~\cite{Amore13B}.

As a result in this limit we have
\begin{equation}
\begin{split}
Z(2) &\approx  \frac{\pi^2}{48} - \frac{5}{32} + \frac{5 \rho -8 \pi }{64 \rho  \log ({r_0})} 
- \frac{7 \rho {r_0}^2}{48 (\rho+4 \pi )} + \dots
\end{split}
\end{equation}

\subsection{Simple harmonic oscillator with an anharmonic perturbation }
\label{sub:sho_anh}

The Schr\"odinger equation in this case is
\begin{equation}
\begin{split}
\left[- \frac{\hbar^2}{2M} \frac{d^2}{dx^2} + \frac{1}{2} M \omega^2 x^2 + g x^4 \right] \Psi_n(x) = E_n \Psi_n(x) \ ,
\end{split}
\end{equation}
and it  can be cast in the dimensionless form
\begin{equation}
\begin{split}
\left[- \frac{1}{2} \frac{d^2}{dy^2} + \frac{1}{2} \ y^2 + \rho y^4 \right] \Phi_n(y) = \tilde{E}_n \Phi_n(y) \ ,
\label{eq_SHO_ANH_adim}
\end{split}
\end{equation}
where $y = x \sqrt{\frac{M \omega}{\hbar}}$ and 
\begin{equation}
\begin{split}
\rho \equiv \frac{\hbar g}{M^2 \omega^3} \hspace{1cm} , \hspace{1cm} \tilde{E}_n \equiv \frac{E_n}{\hbar \omega} \ .
\end{split}
\end{equation}

Observing that the WKB approximation tells us that $\tilde{E}_n \approx n^{4/3}$ for $n \rightarrow \infty$, we conclude that $Z(1)$ is finite. However the first term in  
eq.(\ref{eq_Zeta}) diverges at $s=1$ because of the behavior of the eigenvalues of the simple harmonic oscillator, $\tilde{\epsilon}_n = n +1/2$.

The situation worsens for the first order correction in eq.~(\ref{eq_Zeta}): in this case for $n \gg 1$
the summand behaves as
\begin{equation}
\frac{\langle n | V | n\rangle}{\epsilon_n^{1+s}} \propto \frac{1}{n^{-1+s}}   \ ,
\end{equation}
due to the matrix element
\begin{equation}
\begin{split}
\langle n | y^4 | n \rangle &= \frac{3}{4} \left(2 n^2+2 n+1\right) \ ,
\end{split}
\end{equation}
and the series converges for $s>2$. A similar analysis for the diagonal 
contribution to second order reveals the corresponding series converges for $s>3$.

The origin of these problems lies in the fact that we are not describing correctly the asymptotic
behavior of the spectrum and therefore the expansion of the sum rule breaks down at some
finite order, no matter how large $s$ is. If $s$ is sufficiently large, however, and one sums only the first few orders of the expansion, the sum rule receives most of its contributions
from the low part of the spectrum, that can be well described in the SHO basis. In this case we expect to obtain a good approximation from our perturbative formula~\footnote{An alternative approach would consist of working with an unperturbed basis with eigenvalues that grow faster that $4/3$, such as for a box with hard walls; in this case the expansion is well--defined and
one could  improve the accuracy of the calculation by enlarging the size of the box and including higher order corrections.}.

\subsection{Transforming the Schr\"odinger equation into the Helmholtz equation}
\label{sub:trans}

Consider the time independent Schr\"odinger equation in one dimension
\begin{equation}
\begin{split}
- \frac{\hbar^2}{2m} \frac{d^2\psi}{dx^2} + V(x) \psi(x) = E \psi(x)  \ ,
\end{split}
\label{eq_tise}
\end{equation}
where $V(x)$ is a potential and $\psi(x)$ obeys Dirichlet boundary conditions at $x=\pm L/2$.

Similarly we consider the Helmholtz equation for a heterogeneous system 
\begin{equation}
\begin{split}
- \frac{d^2\phi}{du^2}  = E \Sigma(u) \phi(u)  \ ,
\end{split}
\label{eq_helmht}
\end{equation}
where $\Sigma(u)$ is density and $\phi(u)$ obey Dirichlet boundary conditions at $u = \pm \ell/2$.

We assume
\begin{equation}
\phi(u) =R(u) \ \psi(x(u)) \ ,
\end{equation}
and substitute it inside eq.~(\ref{eq_helmht}) obtaining
\begin{equation}
\begin{split}
- \frac{x'(u)^2}{\Sigma(u)} \frac{d^2\psi}{dx^2}
- \left(\frac{x''(u)}{\Sigma(u)} + \frac{2 R'(u) x'(u)}{R(u)\Sigma(u)}\right) \frac{d\psi}{dx} - 
\frac{R''(u)}{R(u) \Sigma(u)} \psi(x) = E \psi(x) \ .
\end{split}
\label{eq_trans}
\end{equation}

Eq.~(\ref{eq_trans}) takes the form of a time independent Schr\"odinger equation in one dimension provided that
\begin{equation}
\begin{split}
\frac{x'(u)^2}{\Sigma(u)} &= \frac{\hbar^2}{2m} \\
\frac{x''(u)}{\Sigma(u)} &+ \frac{2 R'(u) x'(u)}{R(u)\Sigma(u)} = 0 \ ,
\end{split}
\end{equation}
with a potential
\begin{equation}
V(x) = \frac{R''(u(x))}{R(u(x)) \Sigma(u(x))}  \ .
\end{equation}

The first equation requires
\begin{equation}
x(u) = \frac{\hbar}{\sqrt{2m}}\int \sqrt{\Sigma(u)} du + c_1 \ .
\end{equation}
With the substitution of $x(u)$ in the second equation we obtain 
\begin{equation}
\frac{\Sigma'(u)}{\Sigma(u)} + 4 \frac{R'(u)}{R(u)} = 0 \ ,
\end{equation}
which has the solution
\begin{equation}
R(u) = \frac{c_2}{\Sigma^{1/4}(u)} \ .
\end{equation}

The potential is
\begin{equation}
V(x) =  - \frac{5 \Sigma'(u)^2}{16 \Sigma(u)^3} + \frac{\Sigma''(u)}{4\Sigma(u)^2} \ .
\end{equation}

Notice that the condition $V(x)=0$, corresponding to a free particle trapped in an infinite well, implies a differential equation
for the density of the string with general solution (originally found by Borg in \cite{Borg46})
\begin{equation}
\Sigma(u) = \frac{\beta}{(1+\alpha u)^4} \ .
\end{equation}

As an example, consider the density
\begin{equation}
\Sigma(u) = \frac{\beta}{(1+\alpha u)^2} \ ,
\label{eq_dens1}
\end{equation}
from which
\begin{equation}
\begin{split}
x(u) &= c_1+\frac{\sqrt{\beta } \hbar  \log (\alpha  u+1)}{\sqrt{2} \alpha  \sqrt{M}} \\
R(u) &= c_2 \beta^{-1/4} \sqrt{\alpha  u+1}  \ ,
\end{split}
\end{equation}
and
\begin{equation}
V(x) =  \frac{\alpha^2}{4\beta} \ .
\end{equation}
corresponding to a particle in a box, with a constant potential $\alpha^2/4\beta$.

The sum rule calculated from the eigenvalues of the Schrodinger equation reads
\begin{equation}
\begin{split}
Z^{Sch}(1) &= \sum_{n=1}^\infty \frac{1}{\frac{\hbar^2 \pi^2 n^2}{2 M L^2}+\frac{\alpha^2}{4\beta}} \\ 
&= \frac{\sqrt{2} \sqrt{\beta } L \sqrt{M} }{\alpha 
	\hbar } \coth \left(\frac{\alpha 
	L \sqrt{M}}{\sqrt{2} \sqrt{\beta } \hbar }\right)-\frac{2 \beta }{\alpha ^2}  \ .
\end{split}
\end{equation}

The same sum rule can be calculated using the Helmholtz equation (eq.(11) of Ref.~\cite{Amore13A})
\begin{equation}
\begin{split}
Z^{Helmholtz}(1) &= \int_{-\ell/2}^{\ell/2} \left( \frac{\ell}{4} - \frac{u^2}{\ell}\right)\Sigma(u) du \\
&= \frac{4 \beta}{\alpha^3 \ell} \ {\rm arctanh} \left(\frac{\alpha  \ell}{2}\right)
-\frac{2 \beta }{\alpha ^2} \ .
\end{split}
\end{equation}

The two expressions are seen to be equivalent after relating $L$ to $\ell$ through the equation
\begin{equation}
L = x(\ell/2)-x(-\ell/2) = 
\frac{\sqrt{2} \sqrt{\beta } \hbar}{\alpha  \sqrt{M}} \  {\rm arctanh}\left(\frac{\alpha \ell}{2}\right) \ .
\end{equation}
In a similar way one can calculate higher order sum rules either directly from the eigenvalues of the Schrodinger equation or using the trace formulas discussed  in Ref.~\cite{Amore13A}.

\section{Conclusions}
\label{sec:concl}

We have discussed the calculation of sum rules $Z(s) = \sum_{n=1}^\infty \frac{1}{E_n^s}$ where
$E_n$ are the eigenvalues of the time--independent Schr\"odinger equation in one or more dimensions and $n$ is the set of quantum numbers identifying a given state. The sum rule 
converges for $s > s_0$, where the value of $s_0$  depends both on the potential and on the dimensionality of the problem. 

Extending the method of Refs.~\cite{Amore19B,Amore19C} we have obtained an explicit formula for the sum rule of order $s$ to second order in perturbation theory. We have applied this formula to a simple problem (linear potential in a box) and we have compared the exact results with precise numerical results obtained applying the Rayleigh-Ritz method, reproducing the latter with great accuracy.

For the special case of a infinite box decorated with an impurity at its interior, we have obtained the exact sum rules for the first 
few integer orders exploiting the possibility of obtaining the Green's function for this problem exactly to all orders.  The sum rule 
has been tested numerically for a set of parameters, solving the transcendental equation for the first $2000$ eigenvalues numerically with great 
precision and then completing the series using the asymptotic behavior of the eigenvalues. 

In two dimensions we have considered a disk, with a impurity distributed on a circle, centered at the origin, and we have calculated the
spectral sum rule of order two exactly. For a fixed size of the impurity and letting the strength of the potential change, one observes
that the sum rule has an infinite number of poles which provide the critical couplings where the energy of a state of given angular momentum
vanishes. 

Finally, we have discussed  a different strategy for calculating the sum rules, which is based on the transformation of the one--dimensional Schr\"odinger equation into a Helmholtz equation for an heterogeneous medium.

\section*{Acknowledgements}
The research of P.A. was supported by the Sistema Nacional de Investigadores (M\'exico). 
The author would like to thank Dr. F.M.Fern\'andez for useful comments and suggestions.



\begin{thebibliography}{Bibliography}
\bibitem{Sukumar90} Sukumar, C. V. "Green’s functions and a hierarchy of sum rules for the eigenvalues of confining potentials." American Journal of Physics 58.6 (1990): 561-565.
\bibitem{Crandall96} Crandall, Richard E. "On the quantum zeta function." Journal of Physics A: Mathematical and General 29.21 (1996): 6795.
\bibitem{Weissman79} Weissman, Yitzhak, and Joshua Jortner. "The isotonic oscillator." Physics Letters A 70.3 (1979): 177-179.
\bibitem{Amore19B} Amore, Paolo, "On the calculation of exact sum rules of rational order for quantum billiards" (2019)
\bibitem{Amore19C} Amore, Paolo, "On the calculation of exact sum rules of rational order for quantum billiards (spectrum with a null eigenvalue)" (2019)
\bibitem{Glasser15} Glasser, M. L., and L. M. Nieto. "The energy level structure of a variety of one-dimensional confining potentials and the effects of a local singular perturbation." Canadian Journal of Physics 93.12 (2015): 1588-1596.
\bibitem{Amore13A} Amore, Paolo. "Exact sum rules for inhomogeneous strings." Annals of Physics 338 (2013): 341-360.
\bibitem{Glasser19} Glasser, M. L. "A note on the Exact Green function for a quantum system." Frontiers in Physics 7 (2019): 7.
\bibitem{Duffy} Duffy, Dean G. Green's functions with applications. Chapman and Hall/CRC, 2015.
\bibitem{Mead91} Mead, Lawrence R., and John Godines. "An analytical example of renormalization in two‐dimensional quantum mechanics." American Journal of Physics 59.10 (1991): 935-937.
\bibitem{Tarrach91}Gosdzinsky, P., and Rolf Tarrach. "Learning quantum field theory from elementary quantum mechanics." American Journal of Physics 59.1 (1991): 70-74.
\bibitem{Jackiw91} Jackiw, R. "MAB Beg Memorial Volume." Diverse Topics in Theoretical and Mathematical Physics (1991): 35.
\bibitem{Bender99} Bender, Carl M., and Lawrence R. Mead. "Dimensional expansion for the delta-function potential." European journal of physics 20.2 (1999): 117.
\bibitem{Holstein14} Holstein, Barry R. "Understanding an anomaly." American Journal of Physics 82.6 (2014): 591-596.


\bibitem{Steiner87} Steiner, Frank. "Spectral Sum Rules for the Circular Aharonov‐Bohm Quantum Billiard." Fortschritte der Physik/Progress of Physics 35.1 (1987): 87-114.
\bibitem{Amore13B} Amore, Paolo. "Exact sum rules for inhomogeneous drums." Annals of Physics 336 (2013): 223-244.\bibitem{Borg46} G. Borg, Acta Mathematica 78 (1946) 1–96. doi:10.1007/BF02421600.

\end{thebibliography}
\end{document}